\documentclass[prb,twocolumn,superscriptaddress]{revtex4-2}

\usepackage{graphicx}
\usepackage{epstopdf} 
\usepackage{dcolumn}
\usepackage{bm}
\usepackage{hyperref} 
\usepackage{amsmath}
\hypersetup{
    colorlinks=true,
    citecolor=blue,
    linkcolor=blue,
    filecolor=magenta,      
    urlcolor=blue,
    }

\newcommand{\TUO}{IT4Innovations, V\v{S}B-Technical University of Ostrava, 17 listopadu 2172/15, 708 00 Ostrava, Czech Republic}

\begin{document}

\title{Complex Magnetic Behavior of the Ce sawtooth chains in CeRhSn$_2$.}

\author{P. Opletal$^{1}$, J. Fik\'{a}\v{c}ek$^{1}$, E. Duverger--N\'{e}dellec$^{1,2}$, A. Thamizhavel$^{3}$, Z. Hossain$^{4,5}$, R. Tarasenko$^{6}$, V. Tk\'{a}\v{c}$^{6}$, D. Legut$^{1,7}$, Bin Shen$^{8}$, P. Gegenwart$^{8}$, and J. Custers$^{}$}

\email[Corresponding author: ]{jeroen.custers@matfyz.cuni.cz}

\affiliation{
$^{}$Charles University, Faculty of Mathematics and Physics, Department of Condensed Matter Physics, Ke Karlovu 5, 121 16 Prague 2, Czech Republic.\\
$^{2}$CNRS, ICMCB UMR5026, Bordeaux INP, Bordeaux University, 33600 Pessac, France.\\
$^{3}$DCMP \& MS, Tata Institute of Fundamental Research, Mumbai 400005, India.\\
$^{4}$Department of Physics, Indian Institute of Technology, Kanpur 208016, India.\\
$^{5}$Institute of Low Temperature and Structure Research, Polish Academy of Sciences ul. Ok\'{o}lna 2, 50-422 Wroc\l{}aw, Poland.\\
$^{6}$ Institute of Physics, Faculty of Science, P.J. \v{S}af\'{a}rik University in Ko\v{s}ice, Park Angelinum 9, Ko\v{s}ice, 041, Slovakia \\
$^{7}$ \TUO \\
$^{8}$EP VI, Center for Electronic Correlations and Magnetism, Institute of Physics, University of Augsburg, D-86159 Augsburg, Germany}

\date{\today}

\begin{abstract}
Conflicting reports exist on the ground state of the intermetallic compound CeRhSn$_2$. This can be rooted in the sawtooth-like arrangement of two inequivalent $\text{Ce}$ sites in the unit cell, which suggests potential geometric magnetic frustration. To resolve, we conducted a comprehensive study on high-quality single crystals of CeRhSn$_2$ by means of magnetization ($M$), specific heat ($C_{\mathrm{p}}$/T), and resistivity ($\rho$). The system exhibits strong magnetic anisotropy, confirming the $b$-axis as the easy magnetic axis. We establish three successive transitions, an AFM order at $T_{\mathrm{N}} = 3.65~\text{K}$, a first-order FM order at $T_{\mathrm{C}} = 1.7~\text{K}$ and final transition, at $T = 1.5~\text{K}$. The transition temperatures are highly field-directional dependent: in a magnetic field, the lowest transition is immediately suppressed while $\mathbf{H} \parallel b$ rapidly merges $T_{\mathrm{C}}$ and $T_{\mathrm{N}}$ into a single second-order transition. Conversely, $\mathbf{H} \parallel c$ suppresses the FM order and reduces $T_{\text{N}}$. Additional {\it ab initio} calculations affirm the FM ground state of CeRhSn$_2$. The observation of an enhancement of the Sommerfeld coefficient ($\gamma = 76.5~\text{mJ/mol}\cdot\text{K}^2$) may arise from geometric frustration, but it is most consistently attributed to weak Kondo hybridization as frustration cannot be conclusively established through our data.
\end{abstract}

\vspace{2pc}

\maketitle

\section{Introduction}
\begin{figure}
	\centering
	\includegraphics[width=0.6\columnwidth,bb = 200 391 428 729,clip]{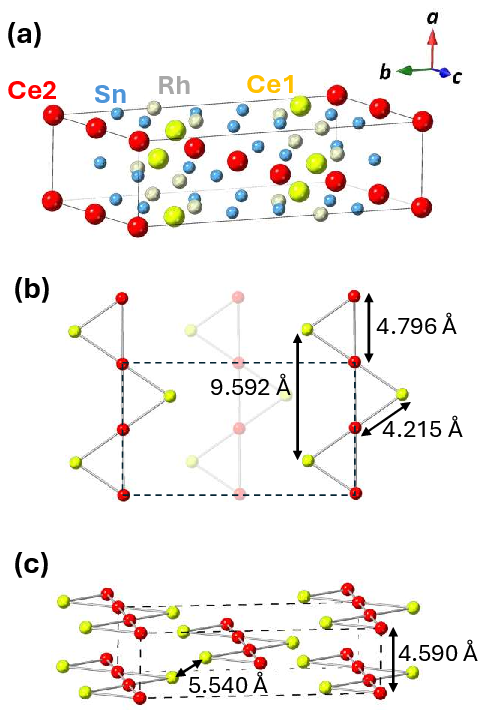}
	\caption{Picture of the crystal structure of CeRhSn$_2$ emphasizing (a) the two inequivalent Ce sites. The view from above of the $bc$ plane (b) shows the magnetic lattice of CeRhSn$_2$. Ce2 (Wyckoff position $4a$) sites align in linear chains along the $c$-axis and link with Ce1 (Wyckoff position $4c$) sites to create a pattern of alternating sawtooth-shaped isosceles triangles. The shortest Ce2--Ce2 distance (c) is along the $a$--axis.}
	\label{Figure1}
\end{figure}

Over the past decades, rare-earth intermetallic compounds containing cerium have commanded considerable scientific interest. These materials exhibit a range of complex phenomena, including valence fluctuations, heavy-fermion behavior, unconventional superconductivity, and various types of magnetic ordering. The origin of these diverse ground states arises from the interaction of $4f$ electrons with the crystal electric field (CEF), as well as the subtle competition between intersite Ruderman–Kittel–Kasuya–Yosida (RKKY) exchange interactions and the on-site Kondo effect. 
When Ce atoms occupy multiple nonequivalent crystallographic sites, the ground state properties become significantly more complex due to the distinct local environments experienced by the $f$ electrons and the presence of additional sublattice interactions. An earlier example is Ce$_2$Sn$_5$ and Ce$_3$Sn$_7$~\cite{Bonnet_JMMM1994}. They are both superstructures of CeSn$_3$ exhibiting two different cerium sites: Ce1 is coordinated by 12 Sn atoms, whereas Ce2 is surrounded by 2 Ce and 10 Sn atoms. 
Despite similar local environments of the Ce atoms, the magnetic ground state of the two compounds differs markedly: Ce$_2$Sn$_5$ experiences strong Kondo interaction evidenced by an enhanced value of $\gamma$ of about 350~mJ/molK$^2$, and orders in a modulated magnetic structure with moments of 1.3$\mu_{\mathrm{B}}$ aligned along the $a$ axis. In contrast, Ce$_3$Sn$_7$ is an antiferromagnet with reduced moments of only 0.36$\mu_{\mathrm{B}}$ parallel to $c$ and a $\gamma$ close to zero. More recent investigated examples are Ce$_{3}$Pd$_{20}$Si$_{6}$~\cite{Nikiforov_JMMM2005,}, CeRuSn~\cite{Mydosh_PRB2011, Fikacek_PRB2012,Feyerherm_PRB2014,Prokes_PRB2015}, Ce$_5$Rh$_4$Sn$_{10}$~\cite{Pathil_PRB1997,Gamza_EURLet2008}, Ce$_3$Pd$_6$Sb$_5$~\cite{Xie_PRB2024} and for instance CeCe$_3$(Pd,Pt)In$_{11}$~\cite{Kratochvilova_SR2015, Prokleska_PRB2015,Fukazawa_PRB2020}. The latter two compounds exhibit antiferromagnetic order coexisting with heavy fermion superconductivity, and have been discussed within a framework in which the two distinct Ce sublattices play separate roles, one driving magnetic ordering and the other facilitating superconductivity.  \\
Against this backdrop, CeRhSn$_2$ emerges as an intriguing case. The compound forms in an orthorhombic structure~ with space group $Cmcm$ (\# 63) and harbors two crystallographically independent cerium sites~\cite{Niepmann_CM1999}. Initial experiments classified the material as nonmagnetic, exhibiting only weak Kondo interactions~\cite{Adroja_PhysicaB1997}. A later study identified a ferro- or ferrimagnetic transition near 4~K~\cite{Niepmann_CM1999}. Subsequent work by Hossain \textit{et al.} uncovered a second magnetic transition around 3~K and, based on low-temperature specific heat fits and magnetoresistance behavior, proposed that the ground state is ferrimagnetic or exhibits canted antiferromagnetic order. The specific heat analysis also yielded a relatively large Sommerfeld coefficient of approximately 300~mJ/molK$^2$, suggesting enhanced quasiparticle masses, characteristic of heavy-fermion systems, or (in part due to) the existence of low-energy magnetic excitations. X-ray photoemission spectroscopy (XPS), in conjunction with local (spin) density approximation (L(S)DA) calculations, revealed that both Ce sites adopt a stable Ce$^{3+}$ configuration, each carrying a moment of $\sim 1\mu_{\mathrm{B}}$. Moreover, the LDA calculations yield an expected Sommerfeld coefficient of only 15.6~mJ/mol·K²~\cite{Gamza_JPCM2009}, significantly lower than the experimentally obtained value. \\
The peculiar arrangement of the Ce ions in CeRhSn$_2$ could be responsible for the observed enhancement in $\gamma$. As illustrated in Fig.\ref{Figure1}, the Ce ions form an equilateral triangular pattern, closely resembling the sawtooth geometry found in, for instance, Mn$_2$Si$X_4$ ($X =$ S, Se)\cite{Mandujano}. Such a geometry is prone to magnetic frustration, which may contribute to the enhanced $\gamma$ coefficient. Alternatively, strong Kondo interactions could underlie the observed heavy-fermion-like behavior, placing CeRhSn$_2$ in proximity to a ferromagnetic quantum critical point (FM QCP). To date, only a limited number of Ce- and Yb-based compounds are known to host a FM QCP~\cite{Brando_RMP2016}. Moreover, it has been proposed that in systems with multiple magnetic sublattices, each sublattice may be characterized by a distinct Kondo scale, potentially giving rise to a complex phase diagram with multiple quantum phase transitions~\cite{Benlagra_PRB2011}.\\
Together, these considerations identify CeRhSn$_2$ as a particularly promising system for further investigation. As a first step, it is essential to resolve the discrepancies in the reported magnetic ordering and to establish a precise characterization of its magnetic properties. In this work, we report on the synthesis of single crystals of CeRhSn$_2$ and a comprehensive study of their low-temperature properties. The magnetic response was examined using dc magnetization, specific heat, and electrical resistivity measurements. Complementary \textit{ab initio} calculations were also performed to independently determine the ground state.

\section{Methodology}
\subsection{Experimental details}

Single crystals of CeRhSn$_2$ were synthesized in India (sample \#1) and subsequently reproduced in Prague (crystal cut in samples \#2–\#5) using the same growth procedure, in order to verify or extend selected experiments. Appropriate amounts of high-purity elements (Ce: 99.95\% and SSE purified ~\cite{Jordan_CP1974,Ford_JLCM1987}, Rh: 99.95\% and Sn: 99.995\%) were melted together in an arc-melting furnace under argon atmosphere. The obtained precursor was turned over and remelted several times to achieve a better homogeneity. Following this, single crystals were grown from the molten precursor using the Czochralski method under an Ar atmosphere. Crystal \#1 was prepared in a tetra-arc furnace, while the second crystal (sample \#2 - \#5) was grown using a tri-arc furnace. In both cases, a tungsten rod was used as a seed rod, and a pulling rate of 12 mm/h was deployed.\\
Subsequently, the crystal structure and chemical composition for both crystals were verified by X-ray Laue method (Laue diffractometer of Photonic Science), powder X-ray-diffraction (XRD, Bruker AXS D8 Advance X-ray diffractometer using the Cu $K_\alpha$ line), and energy dispersive spectrometer (EDS) analysis (scanning electron microscope Tescan Mira I LMH). 
A small piece of size $49\times25\times16 \mu$m$^3$ from crystal \#1 and of size $62\times45\times22 \mu$m$^3$ from sample \#2 was cut off for further investigation by means of X-ray diffraction experiments. On both, single crystal diffraction data were collected at 293~K using an automatic 4--circle Bruker Kappa-Apex II diffractometer equipped with an Apex II CCD detector and a Mo x-rays source with a graphite monochromator. The collection strategy was performed employing Apex II software~\cite{Bruker} and the data were analyzed using CrysAlisPRO software~\cite{CrysAlis}. The cell parameters were defined, and the study of the reciprocal $(hkl)^*$ planes, reconstructed using the unwarp option, was done to evidence systematic absences. Integration of the data was performed with CrysAlisPRO and an empirical absorption correction using spherical harmonics (ABSPACK scaling algorithm) was applied. The structural resolution and refinement were done using Superflip~\cite{Palatinus_JAC2007} and Jana2006~\cite{Petricek_ZK2014} programs. The analysis revealed that the CeRhSn$_2$ single crystals crystallize in an orthorhombic lattice with space group $Cmcm$. The refined unit-cell parameters are $4.5905(10)$\AA{}, $b= 16.9758(5)$\AA{} and $c= 9.5924(3)$\AA{} for crystal \#1, and $a = 4.5913(9)$\AA{}, $b = 16.9824(6)$\AA{}, and $c = 9.5915(3)$\AA{} for sample \#2. These values are in good agreement with previously reported data~\cite{Niepmann_CM1999,Hossain_JPCM2002}.
Inspection of the reconstructed reciprocal planes $(hkl)^*$ reveals a systematic absence condition $h + k = 2n + 1$, consistent with $C$-centering. The refinement of crystal \#1 was performed using 11114 measured reflections, of which 899 were independent with $I > 3\sigma(I)$. A total of 32 parameters were refined. Twinning by a $180^\circ$ rotation about the $(100)$ axis was taken into account, with the minor twin component refined to a volume fraction of 0.10.
For sample \#2, 10591 reflections were measured, yielding 734 independent reflections. In this case, 30 parameters were refined, and no evidence of twinning was observed. The final agreement factors are $R = 2.49$\%)for crystal \#1 and $R = 2.18$\% for crystal \#2. These values demonstrate the good quality of the resolved structure. In addition, it is supported by the theoretical calculations obtaining $a= 4.5971$\AA{}, $b= 17.2251$\AA{}, and $c= 9.6186$\AA{}, see Section \ref{theo} for more details.
\\
Magnetization measurements were conducted on a SQUID magnetometer (MPMS-7, Quantum Design) in the temperature range from 1.8~K to 300~K and in external fields up to 7~T. Subsequent measurements down to 0.4~K were performed using a Quantum Design MPMS-3 SQUID magnetometer with $^3$He option in Ko\v{s}ice and Augsburg. Heat capacity measurements using the relaxation method and resistivity measurements utilizing the standard low-frequency ac four-probe technique were carried out on a commercial physical property measurement system (PPMS-9, Quantum Design) at a $T$--interval of $0.35 < T < 10$~ K.

\subsection{Computational method}
Quantum-mechanical ({\it ab initio}) calculations were performed within density functional theory (DFT)~\cite{DFT1,DFT2} by using the generalized gradient approximation parametrized by Perdew-Burke-Ernzerhof~\cite{PBE} for electron exchange and correlation effects as implemented in the Vienna Atomistic Simulation Package (VASP) code.\cite{VASP} Atoms were represented by the projector-augmented wave pseudopotentials provided by VASP with the following electron configurations 5s$^2$5p$^6$4f$^1$6s$^2$5d$^1$, 5s$^2$5p$^2$ and 5s$^2$4d$^8$ for Ce, Sn, and Rh atoms, respectively. A plane-wave expansion of 500 eV, combined with conjugate gradient energy minimization and quasi-Newton force minimization, was used to
optimize the geometry and atomic positions of the primitive cell (8 formula units). Atomic positions were relaxed until the forces were smaller than $10^{-3}$  eV/\AA, while the total energy was converged down to $10^{-6}$ eV. Brillouin zone was sampled using the {\it k}-point mesh generated by the $\Gamma$-centered Monkhorst-Pack scheme with equidistant length vectors of 100~\cite{VASP}, equivalent to a grid of 22$\times$6$\times$10.

\section{Results}
\subsection{Magnetic properties of CeRhSn$_2$}

\begin{figure}[t]
	\centering
		\includegraphics[width=\columnwidth, angle=0, clip]{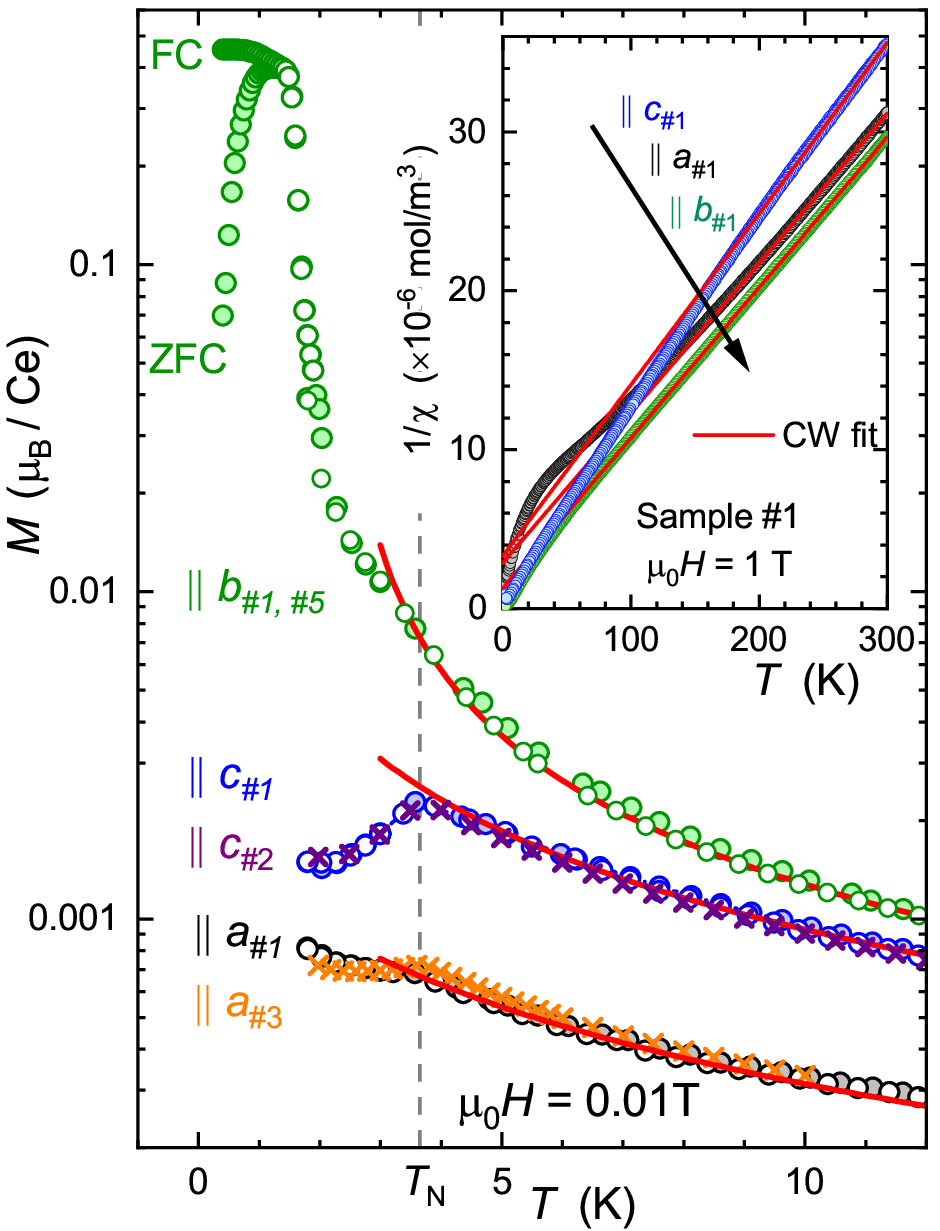}
	\caption{Evolution of the magnetization $M$ at low temperature and in an applied field of $\mu_0H = 0.01$~T. Closed and open symbols denote zero-field-cooled (ZFC) and field-cooled (FC) measurements, respectively, with subscripts indicating sample numbers. The red lines are low-temperature Curie-Weiss fits ($4<T<10$~K). Measurements $T<2$~K ($T\geq 1.8$)~K in the orientation $\mathbf{H} \parallel b$ were carried out on sample \#5 (\#1). The inset plots the inverse of magnetic susceptibility $\chi^{-1} = H/M$ vs temperature $T$ in a field of $\mu_0H = 1$~T along the $a$ axis (black circle), $b$ axis (green circle), and $c$ axis (blue circle).The red lines represent Curie-Weiss fits applied to the data within the temperature range of $200 < T < 300$~K.}
	\label{Figure2}
\end{figure}

\begin{figure}
	\centering
		\includegraphics[width=\columnwidth, angle=0, clip]{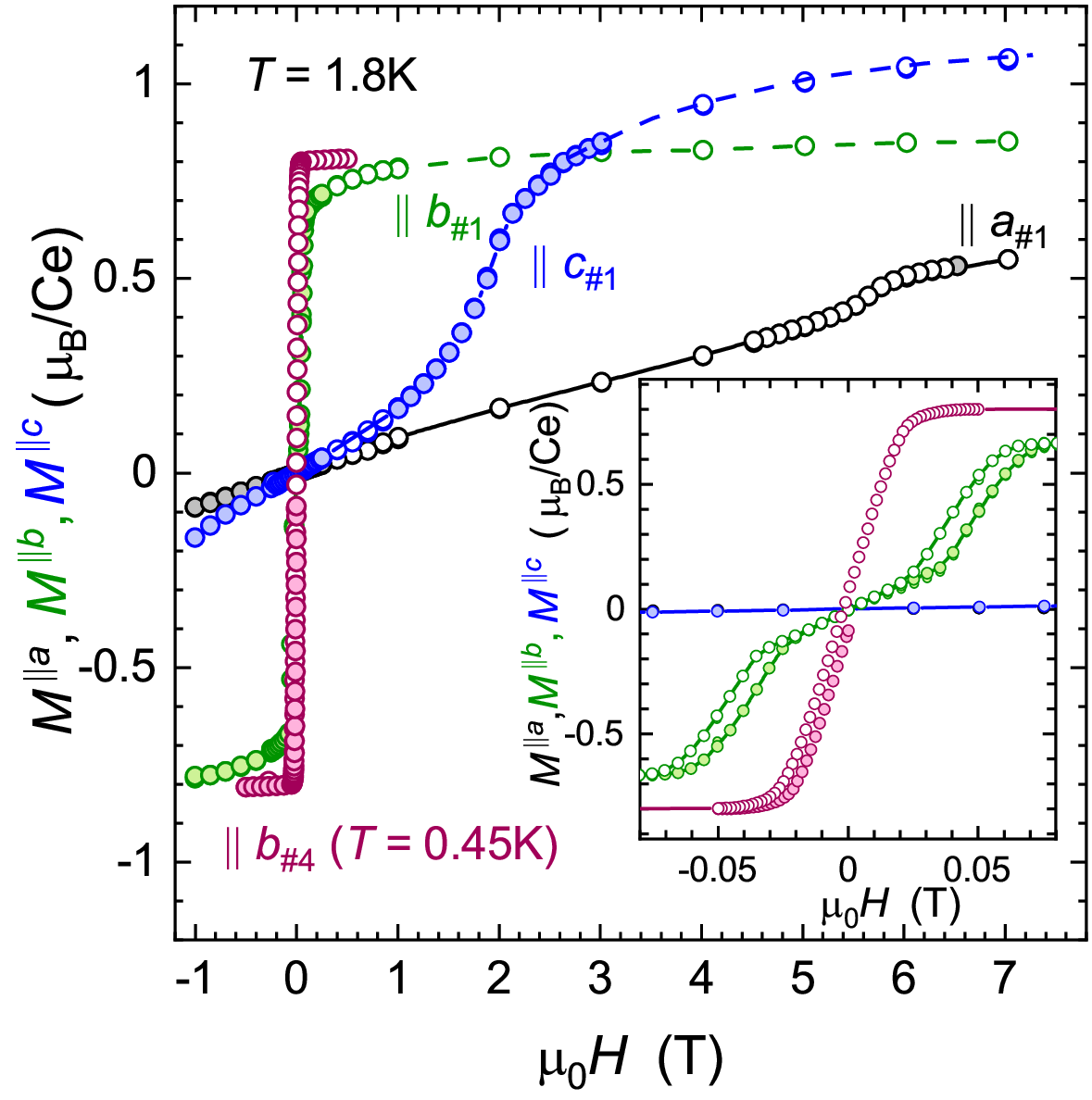}
	\caption{Field dependence of the magnetization for field applied $a$, $b$, and $c$ axis measured at 1.8 K and at $T=0.45$~K ($\parallel b$). The inset shows a zoom-in of the low-field region of $M$ vs $H$. Closed (open) symbols show the data in increasing (decreasing) field. The data $\parallel a$ and $\parallel c$ overlap in low fields. A finite $M$ at zero field persists along $b$ in the $T=0.45$~K measurement.}
	\label{Figure3}
\end{figure}

\begin{figure*}
	\centering
		\includegraphics[width=\textwidth, angle=0, clip]{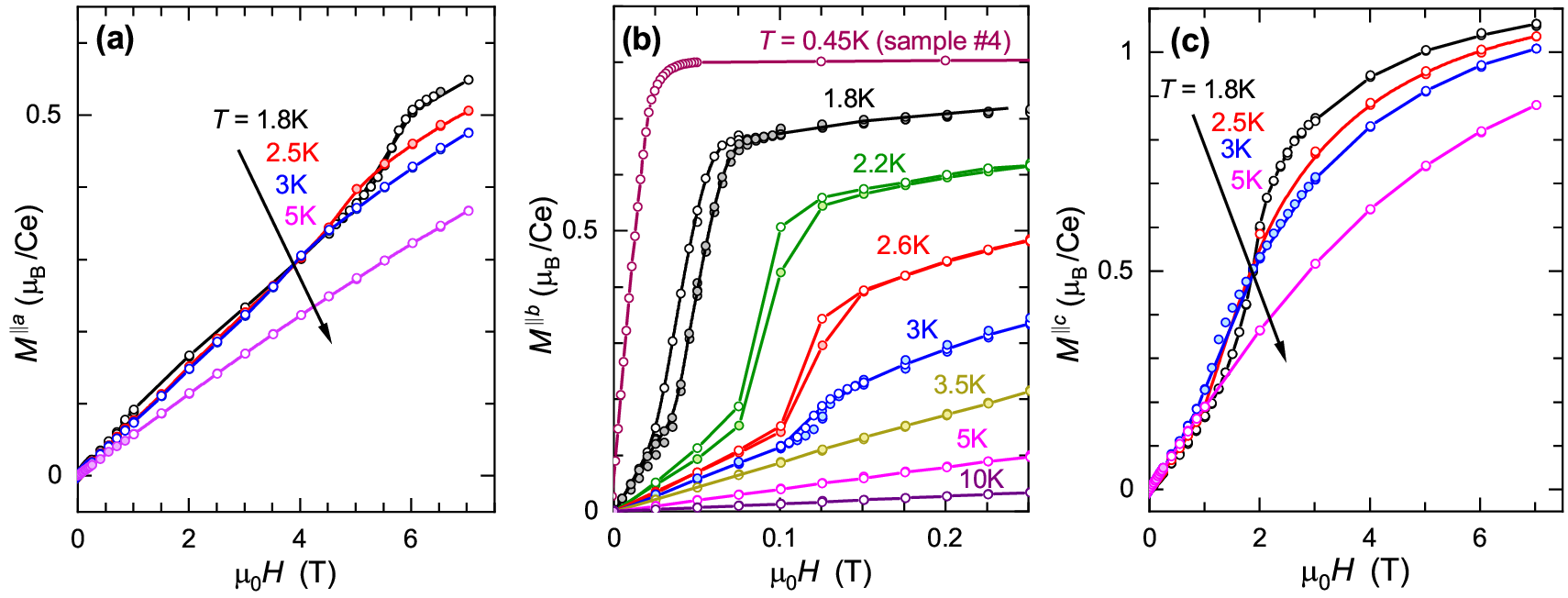}
	\caption{Magnetic field isotherms of CeRhSn$_2$ at various temperatures for $\mathbf{H}$ parallel (a) $a$ axis, (b) $b$ axis, and (c) $c$ axis, respectively. Open (closed) symbols depict ZFC (FC) results.}
	\label{Figure4}
\end{figure*}

The inset of figure~\ref{Figure2} displays the inverse magnetic susceptibility $\chi^{-1}(T) = H/M$ versus the temperature between 1.8 and 300 K measured in a field of 1~T applied along the three main axes. The inverse susceptibility is strongly anisotropic in the entire temperature range. Above 200~K, the  Curie-Weiss (CW) expression $\chi = \frac{N\mu_{\text{B}}^2 \mu_{\text{eff}}^2}{ 3k_{\text{B}}\left(T-\theta_{\text{P}}\right)}$, where the various symbols have their usual meaning~\cite{CW}, provides a good fit to the data giving $\mu_{\text{eff}} = 2.59 \mu_{\text{B}}/{\text{Ce}}$, $\mu_{\text{eff}} = 2.58 \mu_{\text{B}}/{\text{Ce}}$ , and $\mu_{\text{eff}} = 2.44 \mu_{\text{B}}/{\text{Ce}}$
for applied field along the $a,b$, and $c$ direction, respectively, which are in good agreement with the theoretical value of $\mu_{\mathrm{eff}} = 2.54 \mu_{\mathrm{B}}$ for a free Ce$^{3+}$--ion.
Previous reports on polycrystalline specimens~\cite{Niepmann_CM1999,Hossain_JPCM2002} showed that the zero-field-cooled (ZFC) and field-cooled (FC) $M(T)$ curves bifurcate below around $T_{\text{M1}} \approx 4$~K, indicative of a weak ferromagnetic--like transition. Toward lower temperatures, a second transition manifesting through a small cusp in the ZFC cycle emerges at 3~K~\cite{Hossain_JPCM2002}.  In the main panel of Fig.~\ref{Figure1} we present the low temperature ZFC and FC magnetization data ($M^{\parallel a}$ and $M^{\parallel c}$) down to 1.8~K, and $M^{\parallel b}$ down to 0.4~K in $\mathbf{H} = 0.01$~T for various samples. At $T_{\text{N}} = 3.6$~K, a pronounced cusp in $M^{\parallel c}$ signals the onset of antiferromagnetic ordering. Interestingly, the corresponding feature appears only as a faint kink along the $a$-axis and is entirely absent along the $b$-axis, reflecting strong directional dependence in the magnetic response. This is also reflected in the low-$T$ C-W behavior. The fitting of the low-temperature $\chi$ ($4 < T <10$~K) yields $\mu_{\text{eff}} = 1.29\mu_{\text{B}}/{\text{Ce}}$ ($\theta_{\text{P}} = -1.94$~K), $\mu_{\text{eff}} = 2.09 \mu_{\text{B}}/{\text{Ce}}$ ($\theta_{\text{P}} = 2.29$~K), and $\mu_{\text{eff}} = 2.04 \mu_{\text{B}}/{\text{Ce}}$ ($\theta_{\text{P}} = -0.006$~K) for the $a, b$ and $c$ axes, respectively, showing that FM interaction dominates along the $b$ axis. Note that the previously mentioned bifurcation between the ZFC and FC data below $T_{\mathrm{N}}$ is not observed. Instead, for $T<T_{\text{N}}$ the magnetization along this direction increases rapidly and a second transition into a FM order takes place at $T_{\mathrm{C}} = 1.7$~K, marked by a sudden steep increase, reaching a maximum value of $M^{\parallel b} = 0.45 \mu_{\mathrm{B}}/\mathrm{Ce}$ (FC). The splitting of the ZFC and FC curves below $T_{\mathrm{C}}$ ($T_{\mathrm{bf}} \approx 1.5$~K) is in agreement with the expected behavior for a ferromagnet with domain-wall pinning which is enforced by the high magnetocrystalline anisotropy.
\\
To further contrast our data with previous results, we calculate the average susceptibility, defined as $\chi_{\text{avg}} = \frac{1}{3}(\chi^{\parallel a} + \chi^{\parallel b} + \chi^{\parallel c})$, where $\chi^{\parallel a} = M^{\parallel a}/H$ etc. from Fig.~\ref{Figure2} at 5~K yielding a value of $1.47\times10^{-6}$~m$^3$/mol$_{\text{Ce}}$. In comparison, the susceptibility of the polycrystalline sample equals $\chi_{\text{pol}}(5{\text{K}}) = 1.75\times10^{-6}$~m$^3$/mol$_{\text{Ce}}$~\cite{Hossain_JPCM2002}. The discrepancy can be explained due to the preferred orientation of the investigated polycrystalline sample. This is corroborated by isothermal magnetization experiments at 1.8~K (see Fig.~\ref{Figure3}). The average magnetic moment $M_{\text{avg}}$ at the field of 5~T yields 0.74~$\mu_{\text{B}}/\text{Ce}$ which is in excellent agreement with the value of 0.75~$\mu_{\text{B}}/\text{Ce}$ reported in ~\cite{Niepmann_CM1999} but lower than the moment obtained by Hossain {\it et al.}~\cite{Hossain_JPCM2002} being $M = 0.9$~$\mu_{\text{B}}/\text{Ce}$. From our data we conclude that their polycrystalline specimen had a preferred orientation along $c$ direction. \\
More importantly, the isothermal magnetization at $T = 0.45$~K along $\mathbf{H} \parallel b$ displays clear hysteresis, with a remanent magnetization of $0.03\mu_{\mathrm{B}}/\mathrm{Ce}$ at the zero applied field crossing, providing additional evidence for the ferromagnetic nature of the lower transition. The figure also illustrates the strongly anisotropic nature of the intermediate AFM state in CeRhSn$_2$. At $T =1.8$~K, hence slightly above $T_{\mathrm{C}}$, $M^{\parallel b}(H)$ initially displays a linear field dependence. When the field is increased above 0.035~T a sharp metamagnetic transition (MT) occurs, resulting in a magnetization step of 0.6~$\mu_{\text{B}}$. In fields $\mathbf{H} > 0.07$~T, $M^{\parallel b}(H)$ saturates reaching a value of 0.85~$\mu_{\text{B}}$ at 7~T. This is smaller than the $2.14 \mu_{\mathrm{B}}$ value for the Ce$^{3+}$ due to the presence of crystal field effects. A narrow hysteresis occurs when the field is swept down, which closes at $0.015$~ T (see the inset in Fig.~\ref{Figure3}). In the case of $\mathbf{H} \parallel a$, $M(H)$ increases monotonically in field up to 5.3~T where an MT emerges. Unlike for $\mathbf{H} \parallel b$, the MT is broad, more "S" shape-like, and the magnetization step is small from $M^{\parallel a} =$~0.4~$\mu_{\text{B}}/{\text{Ce}}$ at 5.3~T to $M^{\parallel a} =$~0.5~$\mu_{\text{B}}/{\text{Ce}}$ at 6.0~T. In higher fields $M^{\parallel a}$ continues to increase to a value of 0.55~$\mu_{\text{B}}/{\text{Ce}}$ at 7~T. The $M^{\parallel c}$ behavior resembles the former. In low fields the curve initially shows a linear in--$H$ dependence. At $\approx 0.7$~T, an upturn in field sets in followed by a gradual saturation for fields $>2$~T to a value of 1.06~$\mu_{\text{B}}/{\text{Ce}}$ at 7~T. In comparison to the others, the MT is much broader. However, its increase is significant and for $\mu_0H > 2.7$~T the magnetization $M^{\parallel c}$ exceeds $M^{\parallel b}$. In neither of the two field directions, $\parallel a$ and $\parallel c$, a hysteresis in $M(H)$ was found.\\
To characterize the magnetic behavior more comprehensively, we conducted several $M(H)$ measurements at various temperatures below and above $T_{\text{N}}$. These are summarized in Fig.~\ref{Figure4}. The evolution of $M(H)$ for fields $\parallel a$ and $\parallel c$ (Fig.~\ref{Figure4}a and c, respectively) are very similar. For $T < T_{\mathrm{N}}$, which is below the antiferromagnetic ordering temperature, the magnetization initially increases linearly with the increase of magnetic field. As noted earlier, a weak-field-induced metamagnetic transition from AFM to a field-induced polarized state occurs at higher fields, above which magnetization saturates slowly. The transition decreases with increasing temperature. Above 3~K, the MT is no longer observed and $M(H)$ recovers Brillouin-like behavior (Fig.~\ref{Figure4}a and c), consistent with the observation that the system undergoes its initial magnetic ordering near $T_{\mathrm{N}} = 3.6$~K.  In contrast, as the temperature increases, the sharp MT in $M^{\parallel b}(H)$ subsequently broadens and reduces in size, and shifts to higher fields until it disappears for $T \geq 3.5$~K. This behavior is characteristic of FM ordering, and accordingly, we associate the peak position of $dM/dH$ in increasing field (not shown) with the transition from the AFM state into the ferromagnetically ordered state.

\subsection{Specific heat and electrical resistivity of CeRhSn$_2$}

\begin{figure}
	\centering
		\includegraphics[width=\columnwidth, angle=0, clip]{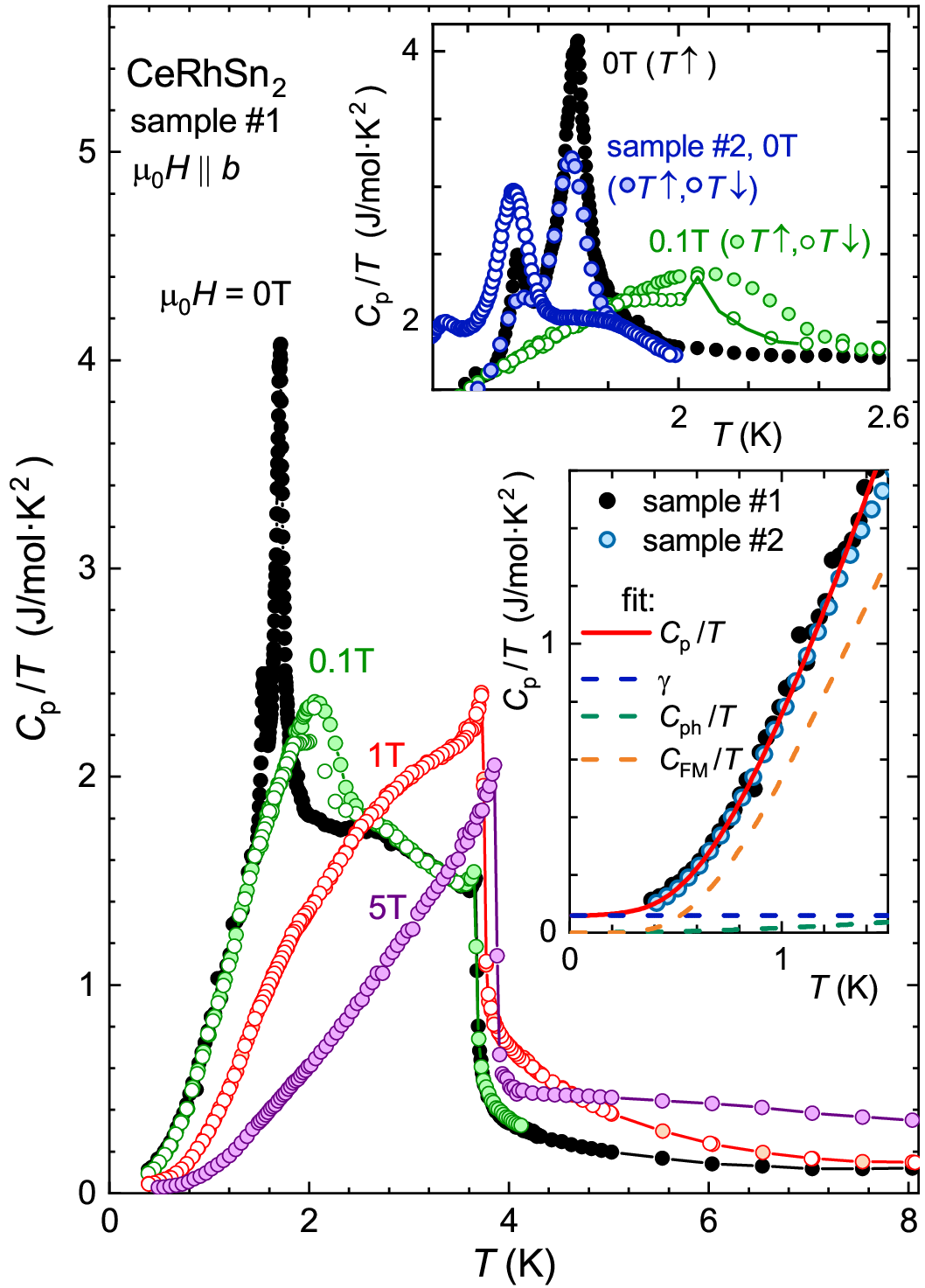}
	\caption{The heat capacity $C_{\mathrm{p}}$ divided by temperature $T$ vs $T$ in various fields applied parallel $b$-axis. Open (closed) symbols is in cooling down (warming up) cycle. The upper inset enlarges the region around the double transition in zero field and from a second sample (\#2). The lower inset presents the fit to the data using Eq.~\ref{eq:HC}.}
	\label{Figure5}
\end{figure}

\begin{figure}
	\centering
		\includegraphics[width=\columnwidth, angle=0, clip]{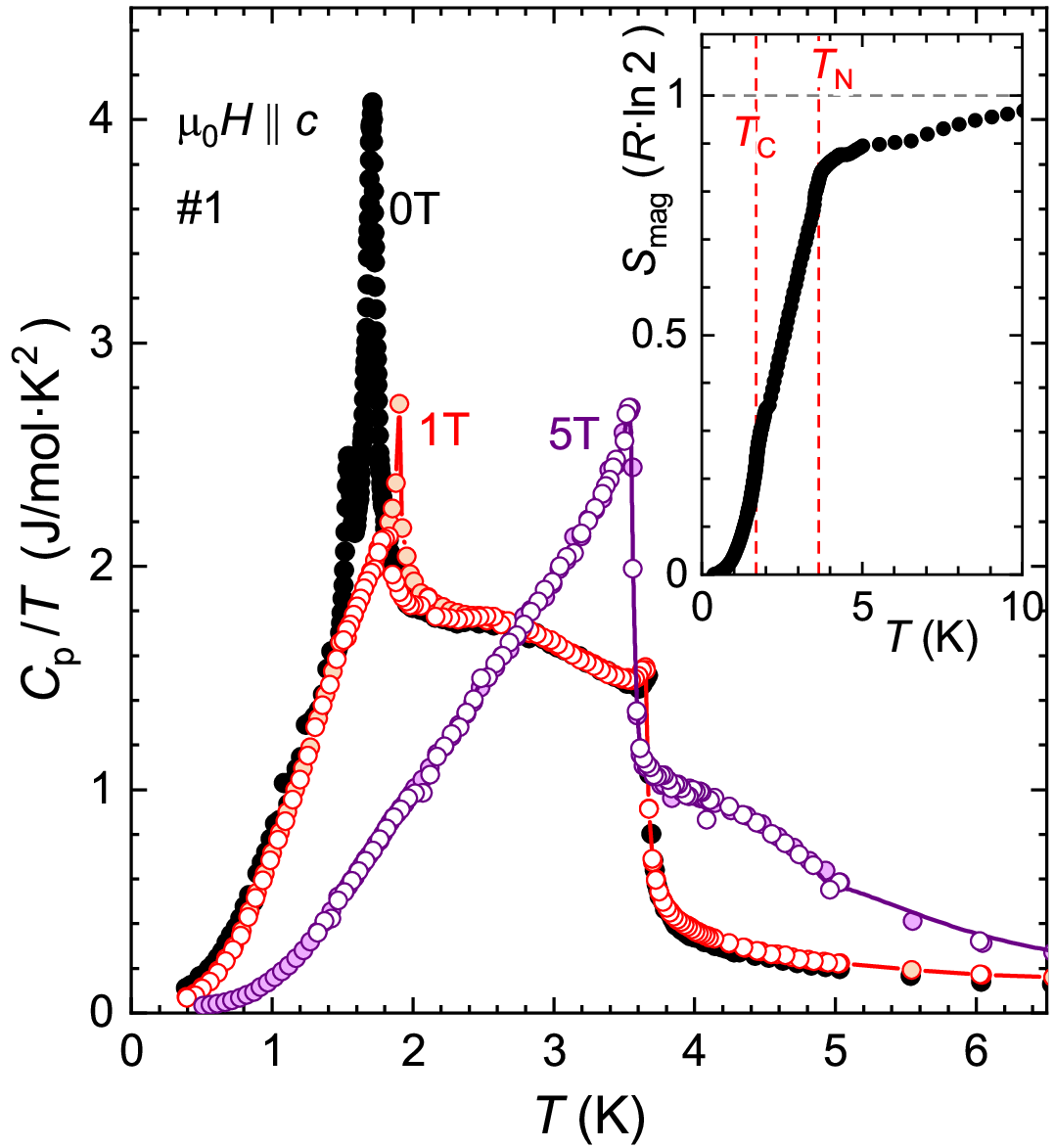}
	\caption{The ratio $C_{\mathrm{p}} / T$ as a function of temperature $T$ of sample \#1, under varying magnetic fields aligned parallel to the $c$-axis, with open (closed) symbols representing cooling down (warming up) cycle. The inset shows the zero field magnetic entropy $S_{\textrm{mag}}$ vs $T$.}
	\label{FigureEntropy}
\end{figure}

Low-temperature heat capacity measurements $C_{\mathrm{p}}/T$ for CeRhSn$_2$ are shown in Fig.~\ref{Figure5} and Fig.~\ref{FigureEntropy} for fields applied parallel to $b$-axis and $c$-axis, respectively. The data is presented without subtraction of a phonon contribution due to the lack of the non-magnetic counterpart LaRhSn$_2$. However, in this temperature range, the phonon contribution is often negligible and the contribution of the $s,p$ and $d$ electrons is small compared to that of the 4$f$ electrons. To elucidate, based on LDA calculations, the Sommerfeld coefficient $\gamma$ of LaRhSn$_2$ is estimated to be of the order of 5~mJ/mol~K$^2$~\cite{Gamza_JPCM2009}. In zero field, $C_{\mathrm{p}}/T$ exhibits a well-defined anomaly at $T_{\mathrm{N}} \approx 3.6\ \text{K}$, consistent with the onset of AFM order inferred from the $M(T)$ data and in agreement with the report by Hossain {\it et al.}~\cite{Hossain_JPCM2002}. However, the low-temperature behavior below $T_{\mathrm{N}}$ differs markedly from previous results. Rather than a single broadened maximum, $C_{\text{p}}/(T)$ increases gradually upon cooling, indicative of residual spin fluctuations within the ordered state.
At lower temperatures, two closely spaced anomalies are resolved. A pronounced, $\lambda$-like peak at $T_{\mathrm{C}} \approx 1.7\ \text{K}$ signals the onset of FM order, followed by a second, smaller anomaly at $T \approx 1.5\ \text{K}$.
Both features exhibit characteristics consistent with weakly first-order behavior and are observed across different samples, indicating their intrinsic nature. Upon further cooling, the heat capacity is rapidly suppressed, reflecting the freezing out of magnetic excitations in the low-temperature ground state.
To resolve the discontinuous character of these transitions, we employed a long heat-pulse technique~\cite{Scheie_JLTP2018} on sample \#1 in the range $1.4 < T < 2\ \text{K}$ (Fig.~\ref{Figure5}, black symbols), while all other data were collected using the standard relaxation method. Clear hysteresis is observed upon temperature cycling, as highlighted in the upper-right inset of Fig.~\ref{Figure5}, corroborating the first-order nature of both transitions. 
Under applied magnetic field, the anomaly at $1.7\ \text{K}$ broadens, shifts to higher temperatures, and evolves toward a more second-order-like character, consistent with the stabilization of FM order. In contrast, the lower-temperature feature at $1.5\ \text{K}$ is progressively suppressed for $\mathbf{H} \parallel b$ and $c$. Its microscopic origin remains unclear but is suggestive of an abrupt spin reconfiguration.\\
Our data further contradict earlier reports~\cite{Hossain_JPCM2002,Gamza_JPCM2009} which suggested a monotonic suppression of $T_{\mathrm{N}}$ under applied magnetic field. Instead, we observe a pronounced anisotropic response: $T_{\mathrm{N}}$ decreases with increasing field for $\mathbf{H}\parallel a$ and $c$, whereas it increases for $\mathbf{H}\parallel b$. At low temperatures $(T < 1.5\ \text{K})$, the specific heat is well described by (see inset of Fig.~\ref{Figure5}):
\begin{equation} \label{eq:HC}
C_{\mathrm{p}}/T = \gamma + \beta T^2 + A\sqrt{T}\exp^{-\frac{\Delta}{T}} .
\end{equation}
Here, $\gamma$ denotes the electronic (Sommerfeld) contribution, $\beta T^2$ represents the phonon contribution, and the last term accounts for the spin-wave contribution of a strongly anisotropic ferromagnet with a gap $\Delta$ in the magnon spectrum~\cite{Slebarski_PRB2004}. The best fit yields $\gamma = 76.5(3)\ \mathrm{mJ/mol\, {\text{K}}^{2}}$, $\beta = 0.67(4)\ \mathrm{mJ/mol\,K^{4}}$, corresponding to a Debye temperature $\Theta_{\mathrm{D}} \approx 226\ \text{K}$, $A = 4.54(7)\ \mathrm{J/mol\,{\text{K}}^{3/2}}$, and a magnon gap $\Delta = 1.92(5)\ \text{K}$, the latter being comparable to $T_{\mathrm{C}}$.
The extracted Sommerfeld coefficient is significantly smaller than the previously reported value of $\sim 300\ \mathrm{mJ/mol\,K^{2}}$, which was obtained from a fit to the rising flank of a broad anomaly~\cite{Hossain_JPCM2002}. Nevertheless, $\gamma$ remains enhanced compared to other Ce-based  ferromagnets such as CeRu$_2$Ge$_2$ ($\gamma \approx 20\ \mathrm{mJ/mol\,K^{2}}$)~\cite{Bohm,Wilhelm} and CePd$_2$Ga$_3$ ($\gamma \approx 9\ \mathrm{mJ/mol\,K^{2}}$)~\cite{Bauer}. Similarly, the previous study using local spin density approximation indicates that the local electronic and exchange effects leads also too low $\gamma$ value ($\gamma \approx 15\ \mathrm{mJ/mol\,K^{2}}$)~\cite{Gamza_JPCM2009}. This is even lower if localization by the Hubbard model is taken into account for Ce $f$-states~\cite{Gamza_JPCM2009}, indicating that in reality many-body electronic effects enhance the states near Fermi level. 
While the obtained value is below that typically observed in strongly correlated Ce-based heavy-fermion systems ($\gamma \gtrsim 100\ \mathrm{mJ/mol\,K^{2}}$), it nevertheless indicates moderate mass enhancement. The inset of Fig.~\ref{FigureEntropy} displays the magnetic entropy $S_{\text{mag}}(T)$ obtained by subtracting the phonon contribution derived from Eq.~\ref{eq:HC} and integrating the resulting $C_{\text{p}}/T$ over temperature. We observe that at the ordering temperature $T_{\mathrm{N}} \approx 3.6\ \text{K}$, the entropy amounts to $S_{\mathrm{mag}} \approx 0.84R\,\ln 2$. The value agrees well with previous results on polycrystalline samples~\cite{Hossain_JPCM2002} but is below the value expected for a well-isolated doublet ground state. The full $R\,\ln 2$ is recovered only at higher temperatures, $T \gtrsim 10\ \text{K}$, consistent with a crystal electric field (CEF) scheme in which the $\mathrm{Ce}^{3+}$ ions possess a well-separated Kramers doublet ground state arising from the splitting of the $J = 5/2$ multiplet. 
The substantial fraction of entropy released at $T_{\mathrm{N}}$ indicates predominantly localized $f$-electron moments. To put this in perspective, known frustrated systems show a significant portion of $S_{\text{mag}}(T)$ being shifted to higher temperatures. For example, in the olivine compound Mn$_2$SiS$_4$, where the Mn ions form a sawtooth lattice~\cite{Nhalil_PRB2019}, the entropy released at the magnetic transition is exceedingly small, amounting to only $3.6\times10^{-5} R\,\ln 6$~\cite{Junod_JMMM1995}. Similarly, in the heavy-fermion metal CePdAl, which crystallizes in the ZrNiAl-type structure and hosts geometrically frustrated Ce moments on a distorted Kagome lattice, the entropy recovered at $T_{\mathrm{N}}$ remains below $0.5 R\,\ln 2$~\cite{Fritsch_JPCS}. In addition, CePdAl exhibits a pronounced maximum in the magnetization above $T_{\mathrm{N}}$ that is not accompanied by a corresponding anomaly in $C_{\mathrm{p}}/T$~\cite{Lucas_PRB2017}. The absence of these features in CeRhSn$_2$ indicates that magnetic frustration plays at most a minor role. In this context, the slightly reduced value of $S_{\mathrm{mag}}$ at $T_{\mathrm{N}}$, together with the moderately enhanced $\gamma$ is more naturally attributed to weak but finite Kondo screening.

\begin{figure}
	\centering
		\includegraphics[width=\columnwidth, angle=0, clip]{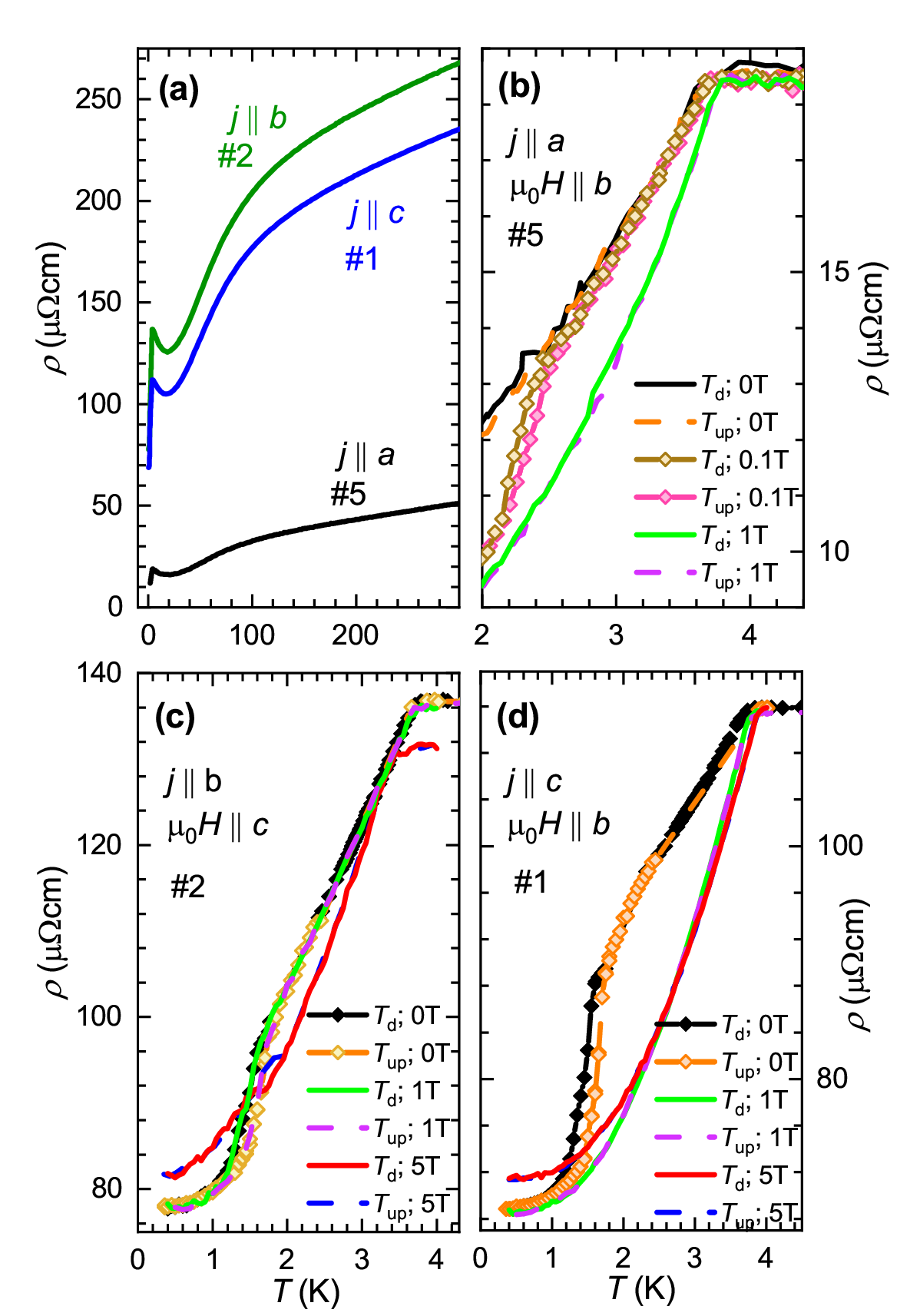}
	\caption{Temperature-dependent electrical resistivity $\rho(T)$. (a) Zero-field data with current density $\mathbf{j}$ applied along the principal axis; measurements were performed on different samples. (b–d) Low-temperature $\rho(T)$ in applied magnetic fields: (b) $\mathbf{H} \parallel b$ with $\mathbf{j} \parallel a$, (c) $\mathbf{H} \parallel c$ with $\mathbf{j} \parallel b$, and (d) $\mathbf{H} \parallel b$ with $\mathbf{j} \parallel c$. Data taken upon cooling and warming are denoted by $T_{\mathrm{d}}$ and $T_{\mathrm{up}}$, respectively.}
	\label{FigureRes}
\end{figure}

\begin{figure}
	\centering
		\includegraphics[width=\columnwidth, angle=0, clip]{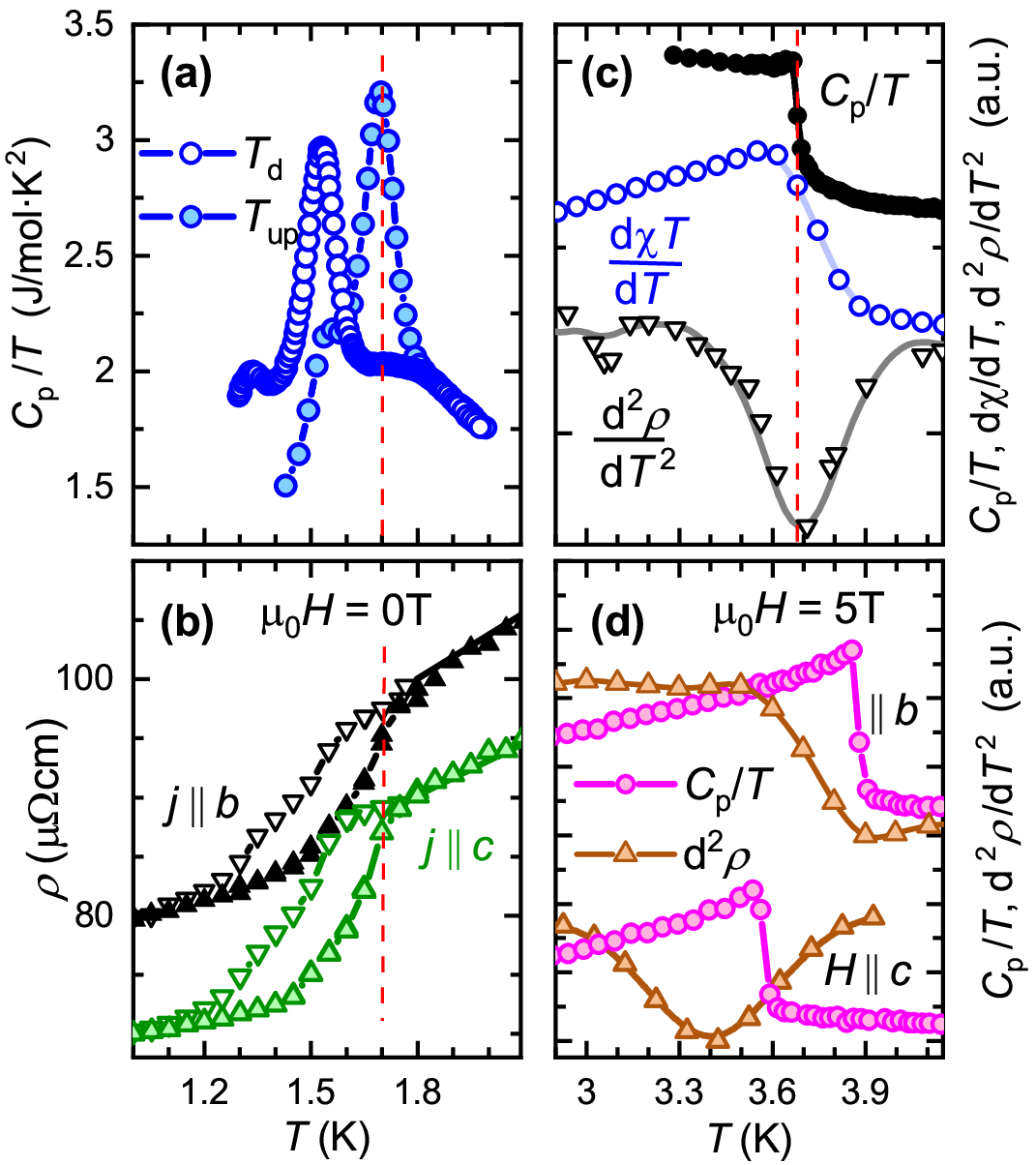}
	\caption{Overview of magnetic transitions: open symbols denote data acquired in the cooling cycle ($T_{\mathrm{d}}$, and closed symbols represent measurements from the warming cycle ($T_{\mathrm{up}}$. (a) $C_{\mathrm{p}}/T$ vs. $T$ in zero field measured using a heat-pulse technique and (b) the respective low-$T$ resistivity for $\mathbf{j} \parallel b$ (black) and $\mathbf{j} \parallel c$ (green). The red dashed line marks $T_{\mathrm{C}}$ obtained from $\chi(T)$ (see Fig.~2). (c) Shows the definition of the AFM transition temperature $T_{\mathrm{N}}$ from $C_{\mathrm{p}}/T$ (zero field), $d\chi T/dT$ (0.01~T), and $d^2\rho/dT^2$ (zero field, $\mathbf{j} \parallel b$) and (d) in a field of $\mu_0H = 5$~T (no susceptibility data available).}
	\label{FigureDetail}
\end{figure}

Figure~\ref{FigureRes}a displays the temperature dependence of the electrical resistivity $\rho(T)$, measured in zero magnetic field with the current $\mathbf{j}$ applied along the principal axes. The data reveal a pronounced anisotropy in the resistivity values with $\rho(T)$ being highest for $\mathbf{j}\parallel b$ (sample \#2), and roughly five times lower when $\mathbf{j}\parallel a$ (sample \#2). Despite this magnitude difference, the temperature evolution of $\rho (T)$ is qualitatively similar across all directions. Upon cooling, the resistivity monotonically decreases, reaching a minimum near 17~K for $\mathbf{j}\parallel b$ and $c$, and around 22~K for $\mathbf{j}\parallel a$. This temperature behavior mimics the previously reported data on polycrystalline samples which also shows a resistance minimum around 20~K~\cite{Adroja_PhysicaB1997,Hossain_JPCM2002} . At lower temperatures, $\rho (T)$ exhibits an upturn followed by a sharp drop, signaling the onset of magnetic ordering at $T_{\mathrm{N}}$. The low-$T$ resistivity regime is shown in more detail in figures~\ref{FigureRes}b, c, and d. Below $T_{\text{N}}$, a second drop in resistivity appears, marking the onset of the FM ordering. This transition displays a clear hysteresis, corroborating the first-order character as previously identified in the $C_{\text{p}}/T$ measurements. However, the third transition, which appears as a subtle peak in $C_{\mathrm{p}}/T$, seemingly lacks a distinct counterpart in the resistivity data. In Figs.~\ref{FigureDetail}a and b, we directly compare the experimental data sets near the ferromagnetic transition. The closed (open) symbols denote the data collected during increasing (decreasing) temperature sweeps. Clearly, the FM anomaly in $C_{\text{p}}/T$ coincides with the drop in $\rho(T)$ in increasing and decreasing temperature runs, while the subsequent small peak in $C_{\text{p}}/T$ may be associated with a faint kink-like irregularity in $\rho(T)$. 

The resistivity response under applied magnetic field is consistent with the behavior observed in the specific heat data. For $\mathbf{H} \parallel b$ (Fig.~\ref{FigureRes}b and d), a small field of $\mu_0 H = 0.1~\text{T}$ strongly enhances the ferromagnetic (FM) transition, shifting $T_{\text{C}}$ from $1.55~\text{K}$ (zero-field, decreasing temperature) up to $2.3~\text{K}$. Concurrently, the kink-like structure in $\rho(T)$, which correlated with the lowest transition observed in $C_{\text{p}}/T$, vanishes. In contrast, the antiferromagnetic (AFM) ordering temperature at $T_{\text{N}} = 3.65~\text{K}$ remains robust and largely unaffected by $0.1~\text{T}$. In a field of $1~\text{T}$ the FM and AFM transitions are merged, as clearly seen in (Fig.~\ref{FigureRes}b and d) and the thermal hysteresis in the resistivity sweep has disappeared. Furthemore, the upper AFM order is slightly pushed to a higher temperature. Conversely, the transitions are much more robust against a field applied along the $c$-axis. For $\mathbf{H} \parallel c$ (Fig.~\ref{FigureRes}c), a field of $1~\text{T}$ does not significantly affect the resistivity, with the $1~\text{T}$ data collapsing on top of the zero-field data within the measurement resolution. No kink-like anomaly below $T_{\text{C}}$ is observed along this current direction. Even at a high field of $\mu_0 H = 5~\text{T}$, we still detect a small thermal hysteresis in $\rho(T)$ around $1.9~\text{K}$. This hysteresis suggests that the FM transition has not fully merged with $T_{\text{N}}$, even though no corresponding anomaly is detected in the $C_{\text{p}}/T$ data at this field. Moreover, the AFM ordering temperature is noticeably reduced at $5~\text{T}$. The field dependence of the magnetic transitions is further detailed in Fig.~\ref{FigureDetail}c and d, where we plot the zero-field and $\mu_0 H = 5~\text{T}$ data for $C_{\text{p}}/T$, the temperature derivative of the magnetic specific heat proxy $d(\chi T)/dT$ ($\mathbf{H} \parallel c$ in $0.01~\text{T}$), and the second derivative of resistivity $d^2\rho/dT^2$. While the function $d(\chi T)/dT$ is generally expected to resemble the $C_{\text{p}}/T$ data near the critical region for simple systems \cite{Fisher1962}, the observed deviation here suggests that we are not dealing with a simple antiferromagnetic transition, highlighting the complexity of the ordering. Crucially, Fig.~\ref{FigureRes}d clearly illustrates that the AFM ordering temperature shifts in opposite directions depending on whether $\mathbf{H}$ is applied along $b$ or $c$ as discussed previously. Furthermore, the discrepancy observed at $\mu_0 H = 5~\text{T}$ ($\mathbf{H} \parallel c$), where the $C_{\text{p}}/T$ anomaly appears at a higher temperature than the minimum in $d^2\rho/dT^2$, suggests a potential influence from sample misalignment. This mismatch could be attributed to a slight tilt of the measured platform (e.g., the heat capacity sample) towards the highly responsive $b$-direction, leading to an apparent shift due to the large magnetic anisotropy.

\subsection{Theoretical results \label{theo}}

\begin{table*}[t]
\begin{ruledtabular}
	\begin{tabular}{cccccccccc}
		& \multicolumn{3}{c}{\textrm{lattice}} & & \multicolumn{4}{c}{\textrm{atomic position}} & \\
		method      & $a$ & $b$ & $c$ & & Ce-$4c$ & Sn-$8f$ & Sn1-$4c$ & Sn2-$4c$ & Rh-$8f$  \\
        \hline
		exp. & 4.5905 & 16.9758  & 9.5924 & & 0.20418 &0.69789/0.58981 & 0.57350 & 0.40848 & 0.64684/0.49519 \\
		theory       & 4.5971  & 17.2251   & 9.6186 & & 0.20446 &  0.698915/0.58935 & 0.57435 & 0.40995& 0.646401/0.49418\\
	\end{tabular}
	\caption{The lattice parameters in \AA\, as well as internal degrees of freedom of atomic positions for various Wyckoff positions of constituent elements. Two numbers (y/z) denote the internal degree of freedom along $y$ and $z$ directions.}
	\label{tab-stru}
\end{ruledtabular}
\end{table*}

The initial spin-polarized calculations were performed, leading to a ferromagnetic ground state. The lattice parameters and internal degrees of freedom of certain atoms are summarized in Table \ref{tab-stru} and compared with experimental data. In general, we obtain very good agreement regarding the lattice parameters (the volume is only 2\% larger than the experimental one) as well as the atomic positions of the optimized internal degrees of freedom; see Table \ref{tab-stru}. The ground state ferromagnetic order exhibits two different spin moments on Ce atoms according to their Wyckoff positions, namely the 4$c$ and 4$a$ amount to 0.4$\mu_{\text{B}}$ and to 0.9$\mu_{\text{B}}$, respectively. The results coincide with the magnetization presented in Figure \ref{Figure3} in an interesting way; if the field is along $b$-axis ($c$-axis), it saturates close to the value of spin moment of Ce at 4$a$ (4$c$) Wyckoff position, that is Ce2 (Ce1). 
However, these would suggest that the external magnetic field will induce antiferromagnetic order on one or the other Ce-sublattice. Even if the starting magnetic ordering was set up to be antiferromagnetic, the ferrimagnetic order prevails, i.e., the Ce atoms carrying significant magnetic moments do not compensate each other; rather, different magnitudes were found, and such ordering possesses higher energy than the pure ferromagnetic order. Hence, we conclude that the FM state is preferred by means of at least 50meV/formula unit with respect to any antiferromagnetic or ferrimagnetic possible arrangements based on the simplest and standard DFT calculations, whereas more sophisticated approaches, such as DMFT~\cite{Kolorenc} or multiplet calculations, might be necessary to fully address the anisotropic behavior in the external magnetic field along $a$, $b$, and $c$-axis.

\section{Conclusion}

We have presented a comprehensive study of single-crystalline CeRhSn$_2$ that resolves longstanding ambiguities regarding its low-temperature properties and magnetic ground state~\cite{Niepmann_CM1999,Adroja_PhysicaB1997,Hossain_JPCM2002,Gamza_JPCM2009}. Thermodynamic and transport measurements performed on multiple single crystals reproduce the behavior reported for polycrystalline samples, demonstrating that the observed features are intrinsic. Our results establish that CeRhSn$_2$ undergoes three magnetic transitions: an antiferromagnetic transition at $T_{\mathrm{N}} = 3.65$~K, followed by ferromagnetic ordering below $T{\mathrm{C}} = 1.7$~K, and a third anomaly at $T = 1.5$~K. The pronounced hysteresis in specific heat and resistivity at the two lower transitions, together with the analysis of thermodynamic data and magnetization measurements, provides clear evidence for a ferromagnetic ground state, thereby resolving the long-standing controversy. This conclusion is further supported by our \emph{ab initio} calculations.

We further show that previously reported enhanced values of the Sommerfeld coefficient $\gamma$ arose from measurements that did not fully capture the completion of ferromagnetic order. The revised value, $\gamma = 76.5$~mJ/mol\,K$^{2}$, is only moderately enhanced compared to local-density approximation calculation. We attribute this to weak Kondo hybridization.

In addition, we demonstrate that the magnetic behavior of CeRhSn$_2$ is highly anisotropic. In particular, the antiferromagnetic transition at $T_{\mathrm{N}} = 3.65$~K is not observed in magnetization for fields applied along the $b$ axis. Furthermore, magnetic fields along specific crystallographic directions appear to selectively stabilize antiferromagnetic correlations on distinct Ce sublattices,  which may account for discrepancies in earlier studies on polycrystalline samples. Although the crystal structure, with its sawtooth-like arrangement of Ce ions, suggests possible geometric frustration, our results indicate that frustration does not play a dominant role.

Hydrostatic pressure measurements are currently underway to explore the possibility of tuning CeRhSn$_2$ toward a ferromagnetic quantum critical point and to examine whether multiple quantum critical points, as proposed in Ref.~\cite{Benlagra_PRB2011}, are realized. In addition, neutron scattering experiments would be highly desirable to resolve the magnetic structures of the ordered phases and to follow their evolution under applied magnetic fields and pressure. Such studies would provide deeper insight into the interplay of competing magnetic interactions in systems with multiple magnetic sublattices.

\section*{Acknowledgments}
J.C. thanks J. Schnack for fruitful discussion. This work was supported by the INTEREXCELLENCE - INTERCOST project no. LUC24139 of Czech Ministry of Education, Youth and Sports (M\v{S}MT) within the framework of the COST Action SUPERQUMAP (CA2144). The experiments were carried out at the Materials Growth and Measurement Laboratory MGML (\url{http://mgml.eu}), which is supported by the Czech Research Infrastructure program (Project No. LM2023065). Z.H. acknowledges financial support from the Polish National Agency for Academic Exchange under the Ulam fellowship. 
D.L. acknowledges the Czech Science Foundation grant No. 25-16339S and the project e-INFRA CZ (ID:90254) for access to computational resources by the Ministry of Education, Youth and Sports of the Czech Republic. Work at the University of Augsburg was supported by the Bavarian-Czech Academic Agency (project no. BTHA-JC-2024-15) and by the German Research Foundation (DFG) through TRR360 (project no. 492547816). The work acknowledges project QM4ST No. CZ.02.01.01/00/22\_008/0004572 funded by MEYS of the Czech Republic.

\end{document}